\documentclass[3p,times,twocolumn]{elsarticle}

\usepackage{ecrc}

\volume{00}

\firstpage{1}

\journalname{Nuclear Physics B Proceedings Supplement}

\runauth{}

\jid{nuphbp}

\jnltitlelogo{Nuclear Physics B Proceedings Supplement}

\usepackage{amssymb}

\biboptions{sort&compress}

\usepackage[figuresright]{rotating}

\usepackage{xspace}
\newboolean{uprightparticles}
\setboolean{uprightparticles}{false} 
\usepackage{amssymb}
\usepackage{amsfonts}
\usepackage{upgreek}

\ifthenelse{\boolean{uprightparticles}}%
{

 \def\Ppi         {\ensuremath{\uppi}\xspace}

 \def\PDelta      {\ensuremath{\Delta}\xspace}                 
 \def\PXi      {\ensuremath{\Xi}\xspace}                 
 \def\PLambda      {\ensuremath{\Lambda}\xspace}                 
 \def\PSigma      {\ensuremath{\Sigma}\xspace}                 
 \def\POmega      {\ensuremath{\Omega}\xspace}                 
 \def\PUpsilon      {\ensuremath{\Upsilon}\xspace}

 \def\PB      {\ensuremath{\mathrm{B}}\xspace}                 
                  
 \def\PD      {\ensuremath{\mathrm{D}}\xspace}

 \def\PK      {\ensuremath{\mathrm{K}}\xspace}

 \def\Pb      {\ensuremath{\mathrm{b}}\xspace}

 \def\Pi      {\ensuremath{\mathrm{i}}\xspace}

 \def\Ps      {\ensuremath{\mathrm{s}}\xspace}

}
{

 \def\Ppi         {\ensuremath{\pi}\xspace}

 \mathchardef\PDelta="7101
 \mathchardef\PXi="7104
 \mathchardef\PLambda="7103
 \mathchardef\PSigma="7106
 \mathchardef\POmega="710A
 \mathchardef\PUpsilon="7107
                  
 \def\PB      {\ensuremath{B}\xspace}                 
                  
 \def\PD      {\ensuremath{D}\xspace}

 \def\PK      {\ensuremath{K}\xspace}

 \def\Pb      {\ensuremath{b}\xspace}

 \def\Pi      {\ensuremath{i}\xspace}

 \def\Ps      {\ensuremath{s}\xspace}

}

\def\squark    {{\ensuremath{\Ps}}\xspace}

\def\bquark    {{\ensuremath{\Pb}}\xspace}

\def\pion   {{\ensuremath{\Ppi}}\xspace}

\def\pip    {{\ensuremath{\pion^+}}\xspace}
\def\pim    {{\ensuremath{\pion^-}}\xspace}

\def\pimp   {{\ensuremath{\pion^\mp}}\xspace}

\def\kaon    {{\ensuremath{\PK}}\xspace}
  \def\Kbar    {{\kern 0.2em\overline{\kern -0.2em \PK}{}}\xspace}

\def\Kp      {{\ensuremath{\kaon^+}}\xspace}
\def\Km      {{\ensuremath{\kaon^-}}\xspace}
\def\Kpm     {{\ensuremath{\kaon^\pm}}\xspace}

\def\KS      {{\ensuremath{\kaon^0_{\rm\scriptscriptstyle S}}}\xspace}

\def\Kstar   {{\ensuremath{\kaon^*}}\xspace}

  \def\Dbar    {{\kern 0.2em\overline{\kern -0.2em \PD}{}}\xspace}

\def\B       {{\ensuremath{\PB}}\xspace}
\def\Bbar    {{\ensuremath{\kern 0.18em\overline{\kern -0.18em \PB}{}}}\xspace}

\def\Bz      {{\ensuremath{\B^0}}\xspace}

\def\Bd      {{\ensuremath{\B^0}}\xspace}
\def\Bs      {{\ensuremath{\B^0_\squark}}\xspace}
\def\Bsb     {{\ensuremath{\Bbar{}^0_\squark}}\xspace}

  \def\Y#1S{\ensuremath{\PUpsilon{(#1S)}}\xspace}

\def\Lbar        {{\ensuremath{\kern 0.1em\overline{\kern -0.1em\PLambda}}}\xspace}

\newcommand{\decay}[2]{\ensuremath{#1\!\to #2}\xspace}        

\def\to                 {\ensuremath{\rightarrow}\xspace}

\def\CP                {{\ensuremath{C\!P}}\xspace}

\newcommand{\dms}{{\ensuremath{\Delta m_{\squark}}}\xspace}

\newcommand{\DGs}{{\ensuremath{\Delta\Gamma_{\squark}}}\xspace}

\def\AT#1     {\ensuremath{A_{\mathrm{T}}^{#1}}\xspace}

\def\C#1      {\ensuremath{\mathcal{C}_{#1}}\xspace}                     
\def\Cp#1     {\ensuremath{\mathcal{C}_{#1}^{'}}\xspace}                 
\def\Ceff#1   {\ensuremath{\mathcal{C}_{#1}^{\mathrm{(eff)}}}\xspace}    
\def\Cpeff#1  {\ensuremath{\mathcal{C}_{#1}^{'\mathrm{(eff)}}}\xspace}   
\def\Ope#1    {\ensuremath{\mathcal{O}_{#1}}\xspace}                     
\def\Opep#1   {\ensuremath{\mathcal{O}_{#1}^{'}}\xspace}

\newcommand{\tev}{\ifthenelse{\boolean{inbibliography}}{\ensuremath{~T\kern -0.05em eV}\xspace}{\ensuremath{\mathrm{\,Te\kern -0.1em V}}}\xspace}
\newcommand{\gev}{\ensuremath{\mathrm{\,Ge\kern -0.1em V}}\xspace}
\newcommand{\mev}{\ensuremath{\mathrm{\,Me\kern -0.1em V}}\xspace}
\newcommand{\kev}{\ensuremath{\mathrm{\,ke\kern -0.1em V}}\xspace}
\newcommand{\ev}{\ensuremath{\mathrm{\,e\kern -0.1em V}}\xspace}
\newcommand{\gevc}{\ensuremath{{\mathrm{\,Ge\kern -0.1em V\!/}c}}\xspace}
\newcommand{\mevc}{\ensuremath{{\mathrm{\,Me\kern -0.1em V\!/}c}}\xspace}
\newcommand{\gevcc}{\ensuremath{{\mathrm{\,Ge\kern -0.1em V\!/}c^2}}\xspace}
\newcommand{\gevgevcccc}{\ensuremath{{\mathrm{\,Ge\kern -0.1em V^2\!/}c^4}}\xspace}
\newcommand{\mevcc}{\ensuremath{{\mathrm{\,Me\kern -0.1em V\!/}c^2}}\xspace}

\def\gsim{{~\raise.15em\hbox{$>$}\kern-.85em
          \lower.35em\hbox{$\sim$}~}\xspace}
\def\lsim{{~\raise.15em\hbox{$<$}\kern-.85em
          \lower.35em\hbox{$\sim$}~}\xspace}

\def\tell1  {TELL1\xspace}
\def\ukl1   {UKL1\xspace}

\newcommand{\ie}{\mbox{\itshape i.e.}\xspace}

\def\BsToKSpipi   {\decay{\Bs}{\KS\pip\pim}}

\usepackage{amsmath}

\usepackage{graphicx}

\begin{document}

\begin{frontmatter}

\dochead{}

\title{Probing \CP violation in \BsToKSpipi decays}

\author{Tim Gershon, Thomas Latham and Rafael Silva Coutinho}

\address{Department of Physics, University of Warwick, Coventry CV4 7AL, United Kingdom}

\begin{abstract}
The three-body charmless hadronic decay \BsToKSpipi provides a number of novel possibilities 
to search for \CP violation effects and test the Standard Model of particle physics. 
These include fits to the Dalitz-plot distributions of the decay-time-integrated final state, decay-time-dependent (but without initial state flavour tagging) 
fits to the Dalitz-plot distribution, as well as full decay-time-dependent and flavour tagged fits. 
The relative sensitivities of these different approaches are investigated.
\end{abstract}


\end{frontmatter}

\section{Introduction}
\label{sec:intro}

The search for a new source of \CP violation in addition to that predicted by the CKM 
matrix~\cite{Cabibbo:1963yz,Kobayashi:1973fv} is among the main goals of current particle physics research.
In the quark sector, a number of important tests have been performed by experiments such as BaBar, 
Belle and LHCb~\cite{Antonelli:2009ws,Brodzicka:2012jm,Bevan:2014iga,LHCb-PAPER-2012-031,HFAG}.
This line of investigation will be continued by Belle~II~\cite{Aushev:2010bq} and the 
upgraded LHCb experiment~\cite{CERN-LHCC-2011-001,CERN-LHCC-2012-007}.

One of the most interesting approaches to search for new sources of \CP violation is by 
studying the decay-time distribution of neutral \B meson decays to hadronic final states 
mediated by the loop (``penguin'') $b \to s$ amplitude.  
As-yet undiscovered particles can contribute in the loops and cause the observables to 
deviate from their expected values in the Standard Model (SM)~\cite{Grossman:1996ke,Fleischer:1996bv,London:1997zk,Ciuchini:1997zp}.
Studies of \Bz decays to $\phi\KS$, $\eta^\prime\KS$, $\KS\KS\KS$ and various other final states have been performed for this reason.  
The latest results are consistent with the SM predictions, but improved measurements are needed to be sensitive to small deviations.

Experience from previous experiments has shown that full decay-time-dependent Dalitz-plot 
analysis of a three-body decay (for example $B^0 \to \KS\pi^+\pi^-$) is more sensitive than 
a ``quasi-two-body'' approach (in this example, considering only the $\KS\rho^0$ contribution).
This is particularly notable in the case that broad resonances contribute, since interference 
causes effects to which quasi-two-body approaches have no sensitivity~\cite{Snyder:1993mx,Charles:1998vf,Latham:2008zs}.
Several methods have been proposed to exploit such interferences in $b \to s$ transitions 
to allow determination of underlying parameters such as the CKM phase $\gamma$ with reduced 
theoretical uncertainty~\cite{Ciuchini:2006kv,Ciuchini:2006st,Gronau:2006qn,Gronau:2007vr,Bediaga:2006jk}.
Full decay-time-dependent Dalitz-plot analyses of $B^0 \to \KS\pi^+\pi^-$~\cite{Aubert:2009me,:2008wwa} 
and $B^0 \to \KS K^+K^-$~\cite{Nakahama:2010nj,Lees:2012kxa} have been performed by BaBar and Belle, 
but similar studies of \Bs meson decays have not yet been possible.

First results from LHCb on decays of the \Bs meson via hadronic $b \to s$ amplitudes have, 
however, recently become available.
Decay-time-dependent analyses of $\Bs \to \Kp\Km$~\cite{LHCb-PAPER-2013-040} and 
$\Bs\to\phi\phi$~\cite{LHCb-PAPER-2013-007} have already been performed.
The first observations of $\Bs\to\KS\Kpm\pimp$ and $\Bs \to \KS\pip\pim$ 
have also been reported~\cite{LHCb-PAPER-2013-042}, including information
on contributing \Kstar resonances~\cite{LHCb-PAPER-2014-043}, suggesting
that it will be possible to study \CP violation in these modes in the
future.

One interesting feature of the \BsToKSpipi decays is that an asymmetry in the time-integrated yields across 
the mirror line of the Dalitz plot is a signature of \CP violation~\cite{Burdman:1991vt,Gardner:2002bb,Gardner:2003su}.
This can be exploited to search for \CP asymmetry with either model-independent or model-dependent approaches.
Another important aspect of the \Bs system, with regard to \CP violation searches, is the non-zero width difference $\DGs$ between the mass eigenstates.
Compared to the situation for \Bd decays, the decay-time distribution receives additional terms that do not vanish when integrated over the initial flavour of the \B meson.
This implies that information about \CP violation parameters can be obtained from analyses 
that do not tag the initial flavour, through so-called effective lifetime measurements~\cite{Dunietz:1995cp,Fleischer:2011cw}.
Although analyses that include flavour tagging information will always be more sensitive, 
this method may still be of interest for analyses based on small event samples, since it is 
difficult to achieve high effective tagging efficiency at hadron collider experiments such as LHCb.

The purpose of this paper is to investigate the comparative sensitivity of different methods to search for \CP violation in \BsToKSpipi decays.
The methods that are considered are 
(i) untagged, decay-time-integrated; 
(ii) untagged, decay-time-dependent; 
(iii) tagged, decay-time-dependent.
Only model-dependent methods are included.
The study is based on a simple toy model for the decays, including
contributions only from $K^*(892)$, $K^*_0(1430)$, $\rho(770)$, and
$f_0(980)$ resonances, implemented with the {\tt Laura++} Dalitz-plot fitting package~\cite{Laura++}.

\section{Formalism}
\label{sec:formalism}

The decay-time distribution for the decays of mesons, initially produced as
\Bsb and \Bs flavour eigenstates, to a final state $f$ can be written~\cite{Dunietz:2000cr}
\begin{equation}
\label{eq:cp_uta:td_cp_bs_asp1}
\begin{array}{lcr}
    \multicolumn{2}{l}{
      \frac{d}{dt} \Gamma_{\Bsb \to f} (t) =      
      \frac{{\cal N}_f \, e^{-t / \tau(\Bs)}}{2\tau(\Bs)}
      \Big[ 
      \cosh\left(\frac{\DGs t}{2}\right) +
			S_f \sin(\dms t) -
		} & \\
    &
    \multicolumn{2}{r}{
      \hspace{1mm} 
      C_f \cos(\dms t) +
      A^{\DGs}_f \sinh\left(\frac{\DGs t}{2}\right)
      \Big]\,,
    } \\
  \end{array}
\end{equation}
and
\begin{equation}
  \label{eq:cp_uta:td_cp_bs_asp2}
  \begin{array}{lcr}
    \multicolumn{2}{l}{
      \frac{d}{dt} \Gamma_{\Bs \to f} (t) =  
      \frac{{\cal N}_f \, e^{-t / \tau(\Bs)}}{2\tau(\Bs)}
      \Big[ 
      \cosh\left(\frac{\DGs t}{2}\right) -
			S_f \sin(\dms t) +
    } & \\
    & 
    \multicolumn{2}{r}{
      \hspace{1mm}
      C_f \cos(\dms t) +
      A^{\DGs}_f \sinh\left(\frac{\DGs t}{2}\right)
      \Big]\,, 
    } \\
  \end{array}
\end{equation}
where the mass and width differences between the light (L) and heavy (H) \Bs physical eigenstates are defined as $\dms = m_{\rm H} - m_{\rm L}$ and $\DGs = \Gamma_{\rm L} - \Gamma_{\rm H}$, and the \Bs lifetime is $\tau(\Bs) = \left(\frac{\Gamma_{\rm L} + \Gamma_{\rm H}}{2}\right)^{-1}$ (units with $\hbar = c = 1$ are used).
The coefficients of the $\sin(\dms t)$, $\cos(\dms t)$ and $\sinh\left(\frac{\DGs t}{2}\right)$  terms are often expressed as 
\begin{equation}
  S_f \;\equiv\; \frac{2\, \Im(\lambda_f)}{1 + \left|\lambda_f\right|^2} \, ,
  C_f \;\equiv\; 
  \frac{1 - \left|\lambda_f\right|^2}{1 + \left|\lambda_f\right|^2} \, ,
  A^{\DGs}_f \;\equiv\; - \frac{2\, \Re(\lambda_f)}{1 + |\lambda_f|^2} \, ,
\end{equation}
where the parameter $\lambda_f$ encodes information about \CP violation and is given by $\lambda_f = \frac{q}{p}\frac{\bar{\cal A}_f}{{\cal A}_f}$ where $\bar{\cal A}_f$ and ${\cal A}_f$ are the amplitudes for \Bsb and \Bs decay to the final state $f$ and $q$ and $p$ define the physical eigenstates in terms of their flavour components
\begin{equation}
  | B^0_{s\,{\rm L}} \rangle = p | \Bs \rangle + q | \Bsb \rangle \,, \qquad 
  | B^0_{s\,{\rm H}} \rangle = p | \Bs \rangle - q | \Bsb \rangle \,, 
\end{equation}
with $| p |^2 + | q |^2 = 1$. Note that, by definition, 
\begin{equation}
  \left( S_f \right)^2 + \left( C_f \right)^2 + \left( A^{\DGs}_f \right)^2 = 1 \, .
\end{equation}
In the remainder of this work, it will be assumed that $|q/p|=1$ (\ie\ absence of \CP violation in mixing).

By requiring that the integral over $t$ from zero to infinity of the sum of Eq.~(\ref{eq:cp_uta:td_cp_bs_asp1}) and Eq.~(\ref{eq:cp_uta:td_cp_bs_asp2}) is equal to $\left| {\cal A}_f \right|^2 + \left| \bar{\cal A}_f \right|^2$, the normalisation factor is found to be 
\begin{equation}
  {\cal N}_f = \left( \left| {\cal A}_f \right|^2 + \left| \bar{\cal A}_f \right|^2 \right) \frac{1-y^2}{1+yA^{\DGs}_f} \, ,
\end{equation}
where $y = \tau(\Bs)\DGs/2$.
The correction involving $y$ is the origin of the difference between branching fractions calculated at $t=0$ or after integration over decay time~\cite{DeBruyn:2012wj}.

The discussion above is appropriate for any final state $f$, including two-body decays.
For multibody decays described by the isobar model~\cite{Fleming:1964zz,Morgan:1968zza,Herndon:1973yn}, 
the total amplitude is obtained from a sum of amplitudes from resonant or nonresonant decay channels,
\begin{equation}
  {\cal A}_f = \sum_{j=1}^{N} c_j F_j(f) \,, \ \ \ 
  \bar{\cal A}_f = \sum_{j=1}^{N} \bar{c}_j F_j(f) \,, \label{eq:dpsum} \\
\end{equation}
where $F_j(f)$ are dynamical amplitudes that contain the lineshape and spin-dependence of the hadronic part 
of the amplitude labelled by $j$ evaluated at the point in phase space given by $f$, and $c_j$ are complex 
coefficients describing the relative magnitude and phase of the different decay channels.
Since the $F_j(f)$ terms describe strong dynamics only, they are \CP conserving.
By contrast, the $c_j$ terms can be \CP violating, which is manifested when $\bar{c}_{j}$ differs from $c_j$ 
in either magnitude or phase -- typically this can occur when the amplitude $j$ has contributions from 
both ``tree'' and ``loop'' (or ``penguin'') amplitudes.

The above discussion makes clear how different forms of \CP violation may be manifest in different types of analysis:
\begin{enumerate}[i.]
\item Untagged, decay-time-integrated Dalitz plot: \\
  In the absence of all forms of \CP violation, there is a symmetry between the mirror line in the $\KS\pip\pim$ phase-space.
  This can be broken, for example, by \CP violation in decay to flavour-specific final states, such as $K^{*\pm}\pi^\mp$, since the \Bs and \Bsb decays populate different regions of the Dalitz plot.
  In general one would expect to find larger asymmetries in some local regions of the phase space, and either model-dependent or model-independent methods could be used to search for such effects.
  A model-dependent fit can determine the $C_f$ parameters of Eq.~\ref{eq:cp_uta:td_cp_bs_asp1} and~\ref{eq:cp_uta:td_cp_bs_asp2}.
\item Untagged, decay-time-dependent Dalitz plot: \\
  The $A^{\DGs}_f$ terms of Eq.~\ref{eq:cp_uta:td_cp_bs_asp1} and~\ref{eq:cp_uta:td_cp_bs_asp2} can be determined, and therefore more information is obtained compared to the decay-time-integrated case.
\item Tagged, decay-time-dependent Dalitz plot: \\
  All terms, including the $S_f$ parameters, can be determined.
  This method therefore provides additional sensitivity to the model parameters, in particular to the relative phase between \Bs and \Bsb decay amplitudes.
\end{enumerate}

This general discussion does not answer the question of how much additional sensitivity is obtained as the analysis is made increasingly more complex.
That will be addressed in the next sections.

\section{Method to generate toy samples}
\label{sec:generation}

Several ensembles of Monte Carlo pseudoexperiments are generated 
to investigate \CP violation effects in \BsToKSpipi decays.
The simulation is performed without any experimental effects, such as background, acceptance, resolution or imperfect flavour tagging.
The toy model contains the $\rho^{0}(770)$, $f_{0}(980)$, $K^{*\pm}(892)$ and $K^{*\pm}_{0}(1430)$ resonances.
All mass terms are described by the relativistic Breit-Wigner (RBW) function, apart from the $K^{*\pm}_{0}(1430)$ lineshape which is modelled by the LASS shape~\cite{Aston:1987ir}.  
The parametrisation of complex coefficients is given by 
\begin{equation}
\overset{(-)}{c_{j}} = (x_{j} \pm \Delta x_{j}) + i (y_{j} \pm \Delta y_{j})\,, 
\end{equation}
where $\Delta x_{j}$ and $\Delta y_{j}$ are \CP-violating parameters.
Table~\ref{tab:1} summarises the baseline model used to generate events, with decay-time distribution given in Eq.~\ref{eq:cp_uta:td_cp_bs_asp1} and~\ref{eq:cp_uta:td_cp_bs_asp2}.
Values of $\tau(\Bs) = 1.517 \ {\rm ps}$, $\Delta m_s = 17.76 \ {\rm ps}^{-1}$ and $y = 0.058$ are used.

\begin{table}[!htb]
  \caption{
    Benchmark parameters for the baseline Dalitz plot model used as input in the generation. 
  }
\begin{center}
\resizebox{7.8cm}{!}{
\begin{tabular}{ccccc}
\hline
Resonance            &            $x_{j}$ & $\Delta x_{j}$ &            $y_{j}$ & $\Delta y_{j}$ \\
\hline 
$\rho^{0}(770)$      &              $1.0$ &          $0.0$ &              $0.0$ &          $0.0$ \\
$f_{0}(980)$         & $0.4 \cos(5\pi/4)$ &          $0.0$ & $0.4 \sin(5\pi/3)$ &          $0.0$ \\
$K^{*\pm}(892)$      &  $1.2 \cos(\pi/3)$ &          $0.0$ &  $1.2 \sin(\pi/3)$ &          $0.0$ \\
$K^{*\pm}_{0}(1430)$ &  $1.7 \cos(\pi/3)$ &          $0.0$ &  $1.7 \sin(\pi/3)$ &          $0.0$ \\
\hline
\end{tabular}
}
\end{center}
\label{tab:1}
\end{table}

In the fit, the $\overset{(-)}{c_{j}}$ coefficients are measured relative to the $\rho^{0}(770)$ resonance contribution.
Each pseudoexperiment is fitted many times with randomised initial values of the parameters in order to find the global minimum of the negative log likelihood function.
Asymmetries are calculated as 
\begin{equation}
{\cal{A}}_{\CP\,j} = \frac{|{\bar{c}}_{j}|^{2} - |c_{j}|^{2}}{|{\bar{c}}_{j}|^{2} + |c_{j}|^{2}} = \frac{-2(x_{j}\Delta x_{j} + y_{j}\Delta y_{j})}{x^{2}_{j} + \Delta x^{2}_{j} + y^{2}_{j} +\Delta y^{2}_j} \, .
\end{equation}
\CP violation can also be manifest in a difference between the phase of the \Bs and \Bsb decay amplitudes, 
\begin{equation}
  \Delta \delta_j = \arg\left(\frac{\bar{c}_j}{c_j}\right)  = \tan^{-1}\left( \frac{ y_j + \Delta y_j }{ x_j + \Delta x_j } \right) - \tan^{-1}\left( \frac{ y_j - \Delta y_j }{ x_j - \Delta x_j } \right)\, .
\end{equation}

The baseline model is modified in various ways to introduce \CP violation.
Interference between the \Bs-\Bsb oscillation and decay amplitudes is incorporated through the \CP violation weak phase $\phi_{s}$.
While the SM predicts $\phi^{\rm SM}_{s} = -2\beta_s \equiv -2 \arg (-V_{ts}V^{*}_{tb}/V_{cs}V^{*}_{cb}) = -0.036 \pm 0.002$ rad, 
contributions from physics beyond the SM could lead to much larger values.
Three different scenarios are generated: $\phi_{s} = 0$, $\phi_{s} = -2\beta_{s}$ and $\phi_{s} = -20\beta_{s}$. 
In addition, \CP violation in the decay of each resonance is examined: 
\CP violation in the magnitude, with
${\cal{A}}_{\CP} = 5\%,\,10\%,\,20\%$ and $50\%$; \CP violation due to the
difference in the relative phase in steps of $\pi/4$ from $0$ to $2\pi$; and \CP violation in both magnitude and phase difference.
Pseudoexperiments are generated with sample size corresponding roughly to the anticipated yields available at LHCb by the end of the LHC Run II (2000 events).
Ensembles with other sample sizes are also generated to test the scaling of the uncertainties.
Only a representative subset of the results obtained are presented here due to space constraints.

\section{Results}
\label{sec:results}

\begin{figure*}[htb]
\centering
\includegraphics*[width=0.38\textwidth]{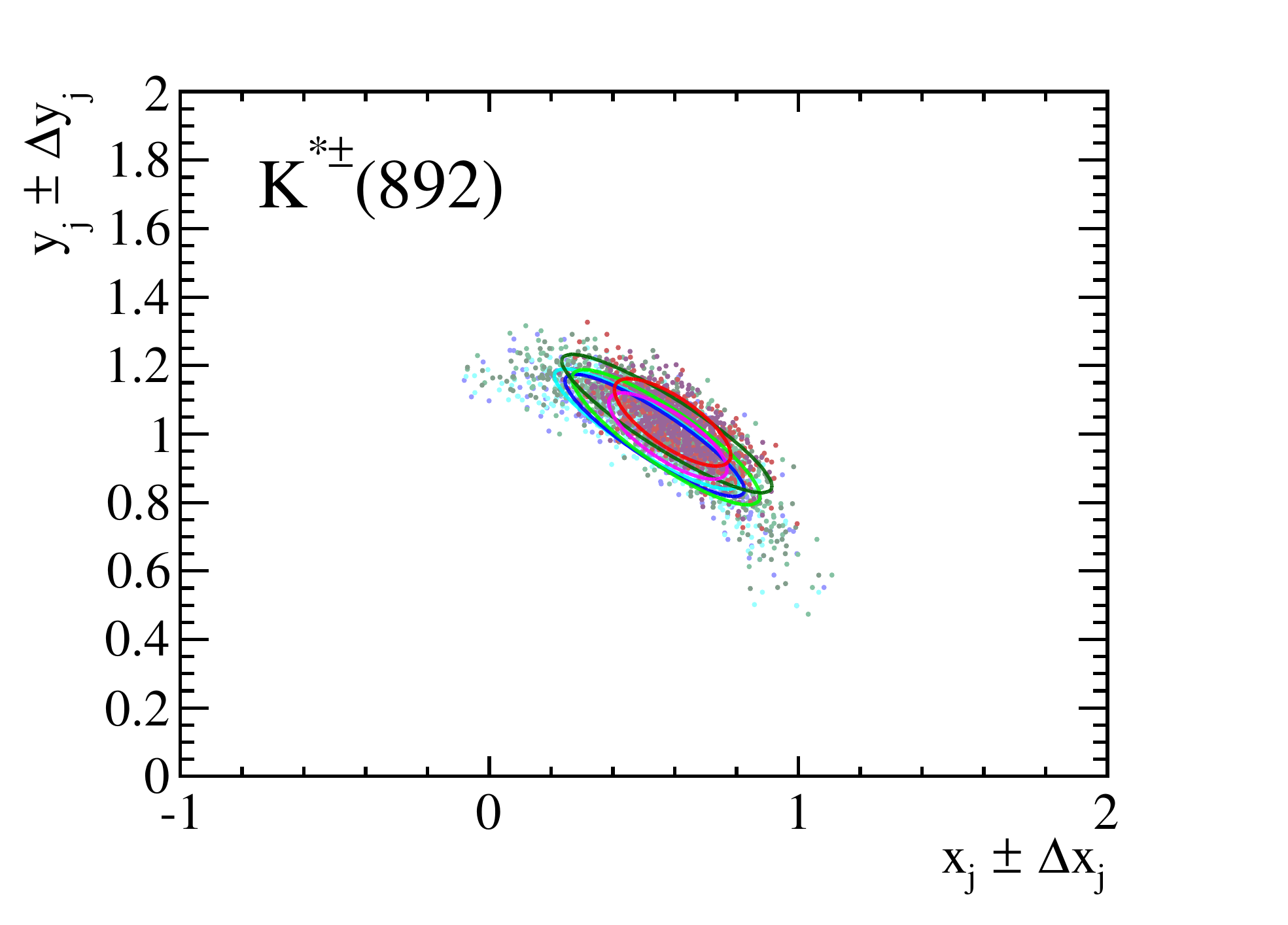}
\includegraphics*[width=0.38\textwidth]{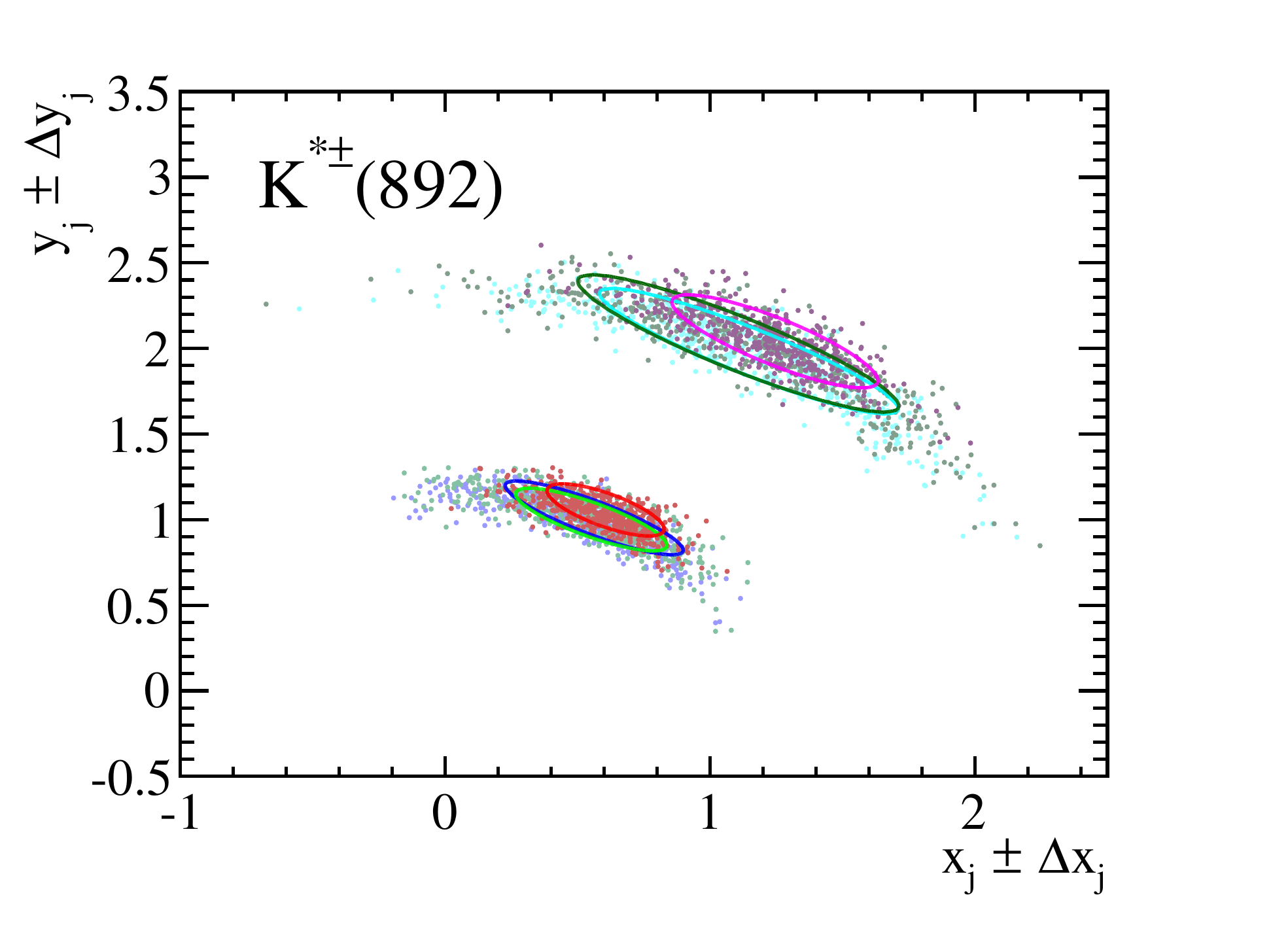}
\includegraphics*[width=0.38\textwidth]{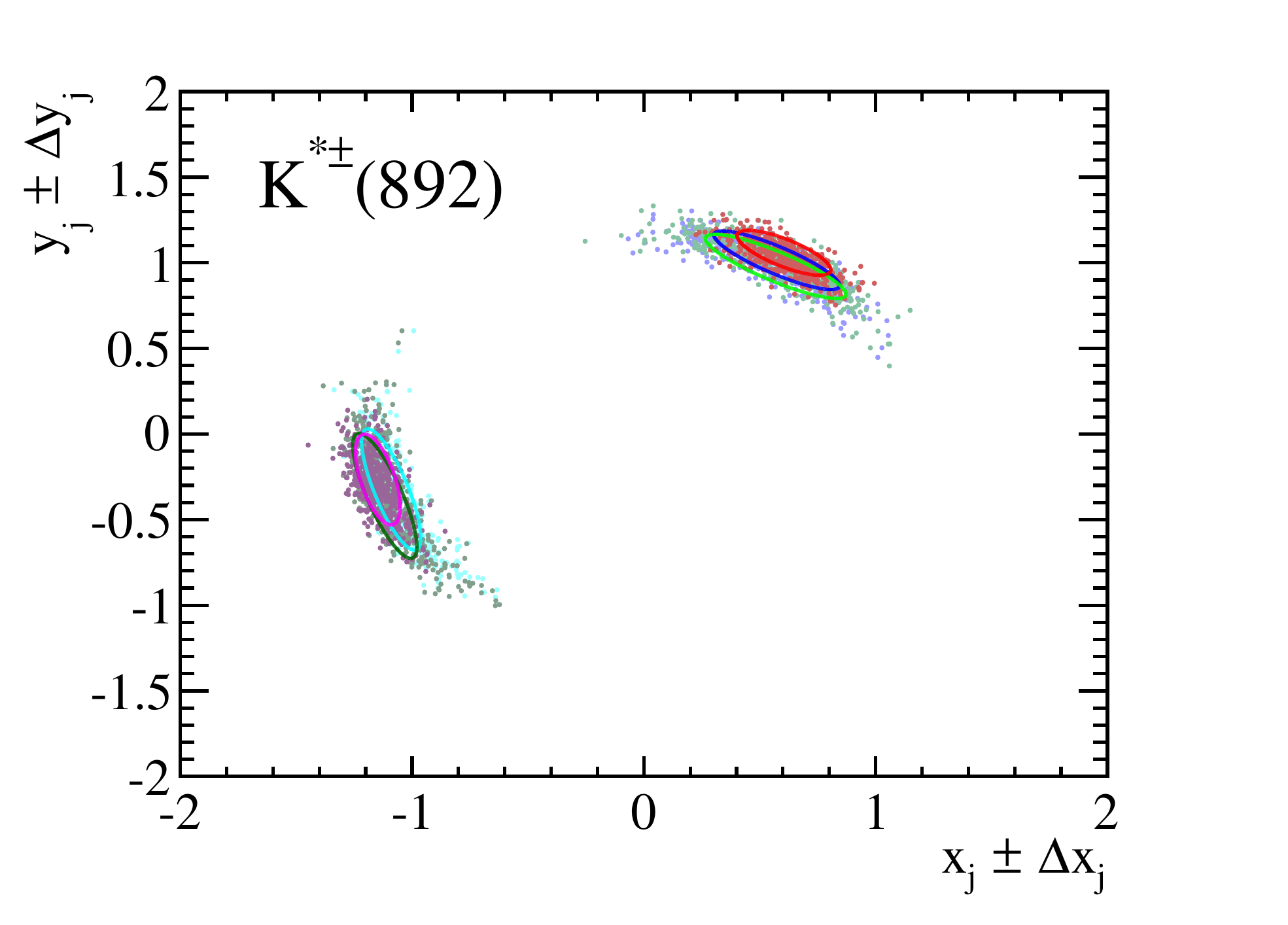}
\includegraphics*[width=0.38\textwidth]{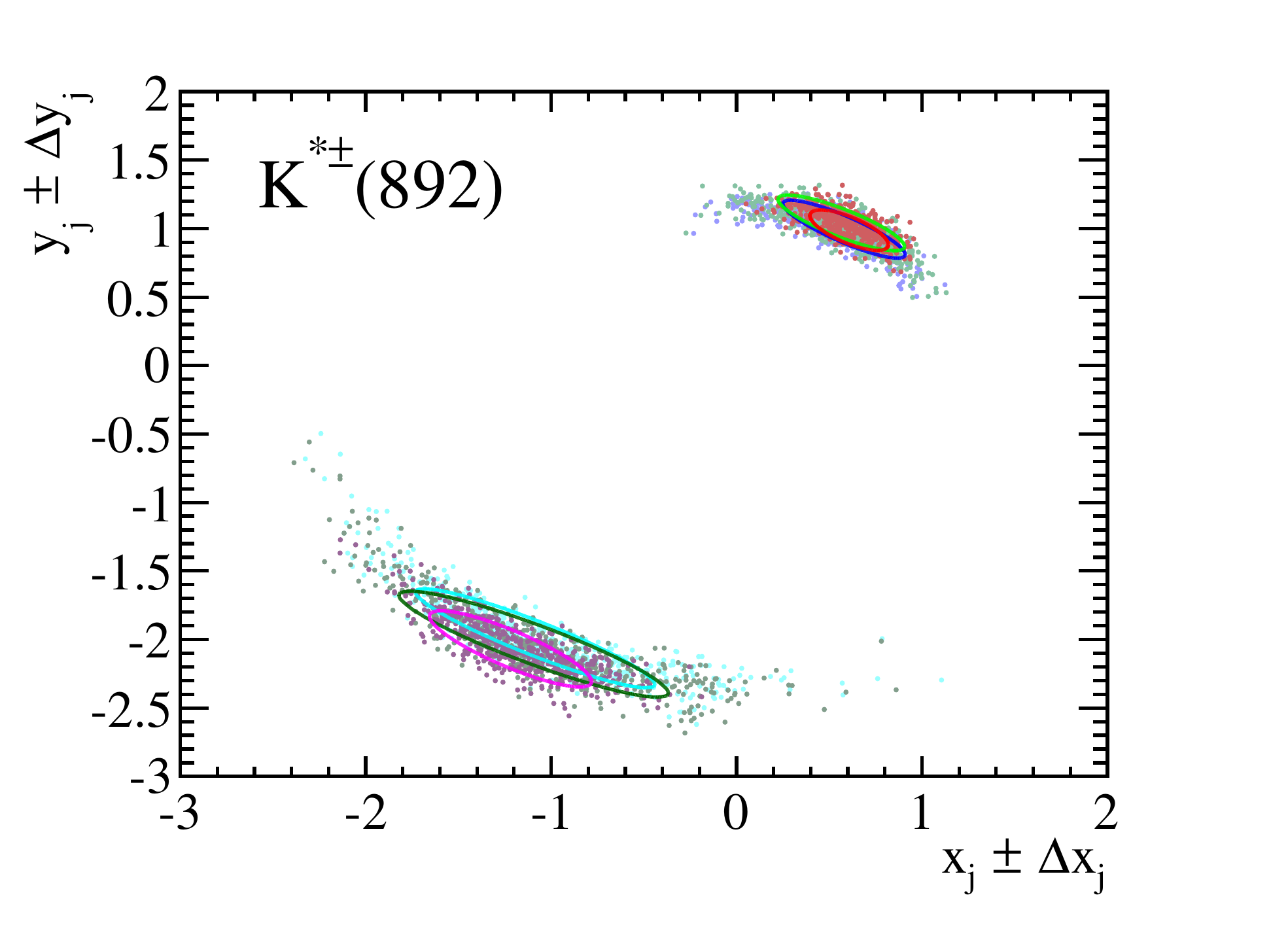}
\caption{
  Fitted values of the $K^{*\pm}(892)$ isobar coefficients plotted in the Argand plane.
  The points are the values determined from individual pseudoexperiments while the ellipses illustrate the mean value and 68\% confidence level contour from the corresponding ensemble.
  The results for $c_j$ ($\bar{c}_j$) are shown for method i in blue (cyan), for method ii in green (dark green) and for method iii in red (magenta).
  All experiments are generated with 2000 signal events and $\phi_{s} = -2\beta_{s}$ and the following scenarios:
  (top left) no \CP violation,
  (top right) ${\cal{A}}_{\CP} = 50\%$, 
  (bottom left) $\Delta \delta = 3\pi/4$ and
  (bottom right) ${\cal{A}}_{\CP} = 50\%$ and $\Delta \delta = \pi$.
}
\label{fig:Kstar_Argand}
\end{figure*}

Figure~\ref{fig:Kstar_Argand} shows the results for various scenarios of \CP violation in the $K^{*\pm}(892)$ amplitude, with yields corresponding to LHC Run I+II.
The fitted values of the isobar coefficients in each pseudoexperiment are represented by the points in the Argand plane, with the ellipses illustrating the central values and $1\,\sigma$ contour boundaries from the ensemble.
The colour schemes for \Bs and \Bsb coefficients are represented respectively by: 
blue and cyan for method i; 
light and dark green for method ii;
and red and magenta for method iii.

It is immediately clear that the magnitude of the amplitude is determined much more precisely than the phase, leading to the arc-like distribution of points.
Table~\ref{tab:Fit_results} compares the precision of the different fitting methods for each of the \CP violation scenarios.
The results indicate that the generated asymmetries are retrieved in all scenarios with good precision and without significant bias.
The untagged methods give statistical uncertainties that are only slightly larger, due to the fact that the $\Kstar$ resonances from the decay of \Bs and \Bsb populate different regions of the Dalitz plot.
In addition, the very similar uncertainties given by the two untagged approaches suggests that the $A^{\DGs}_{f}$ term does not provide a significant amount of extra sensitivity.
Further studies with realistic experimental effects are necessary to determine the exact sensitivities achievable.
An extrapolation of the precision estimated here suggests that such measurements appear to be feasible, albeit with large uncertainty, with the LHCb Run I dataset that is already in hand.

\begin{table*}
\centering
\caption{
  Comparison of the uncertainties on the $K^{*\pm}(892)$ \CP-violating parameters determined using the different fitting methods.
  The results are quoted in terms of the polar co-ordinates $c_j = a_j e^{i\delta_j}$, $\bar{c}_j = \bar{a}_j e^{i\bar{\delta}_j}$.
  The relative uncertainties for method iii are quoted (the central values of the parameters correspond to the values given in Table~\ref{tab:1}, modified according to the \CP violation parameters), together with comparisons of the uncertainties with the different Dalitz plot fit methods.
  The typical uncertainty on the relative precision is $\pm 0.1$.
}
\begin{tabular}{ccccccc@{\hspace{5mm}}cccc@{\hspace{5mm}}cccc}
\hline 
\multicolumn{3}{c}{\CP-violation parameters}     & \multicolumn{4}{c}{$\sigma({\rm method\,iii}) \ (\%)$} & \multicolumn{4}{c}{$\frac{\sigma({\rm method\,ii})}{\sigma({\rm method\,iii})}$} & \multicolumn{4}{c}{$\frac{\sigma({\rm method\,i})}{\sigma({\rm method\,iii})}$}\\
${\cal{A}}_{CP}$ &  $\Delta\delta$ & $\phi_{s}$    &   $a_{j}$ & $\bar{a}_{j}$ & $\delta_{j}$ & $\bar{\delta}_{j}$  &   $a_{j}$ & $\bar{a}_{j}$ & $\delta_{j}$ & $\bar{\delta}_{j}$  &   $a_{j}$ & $\bar{a}_{j}$ & $\delta_{j}$ & $\bar{\delta}_{j}$ \\
\hline \hline
$20\%$ & $0$  & $0$          & $4.6$ & $3.7$  & $12.3$ & $11.6$ & $1.1$ & $1.2$ & $1.8$ & $1.8$ & $1.0$ & $1.0$ & $1.7$ & $1.8$\\
$50\%$ & $0$ & $2\beta_{s}$  & $5.1$ & $3.3$  & $15.0$ & $12.2$ & $1.0$ & $1.1$ & $1.6$ & $1.8$ & $0.9$ & $1.1$ & $1.5$ & $1.8$  \\
$0\%$ & $\pi/4$ & $0$ & $4.2$   & $4.3$ & $12.2$ & $7.7$  & $1.0$  & $1.1$ & $1.8$ & $1.6$ & $1.0$ & $0.9$ & $1.6$ & $1.5$ \\
$0\%$ & $3\pi/4$ & $2\beta_{s}$ & $4.2$ & $4.0$  & $12.4$ & $4.7$  & $1.0$ & $1.2$ & $1.6$ & $1.8$ & $1.0$ & $1.1$ & $1.5$ & $1.8$\\
$5\%$ & $\pi/4$ & $0$         & $4.5$ & $3.9$ & $11.4$ & $8.3$ & $1.0$ & $1.1$ & $1.8$ & $1.4$ & $0.9$ & $1.0$ & $1.8$ & $1.5$ \\
$50\%$ & $\pi$ & $2\beta_{s}$ & $5.2$ & $3.6$ & $14.5$ & $7.1$ & $1.0$ & $1.1$ & $1.6$ & $1.7$ & $1.1$ & $0.9$ & $1.6$ & $1.7$\\
\hline
\end{tabular}
\label{tab:Fit_results}
\end{table*}

\begin{figure*}[htb]
\centering
\includegraphics*[width=0.38\textwidth]{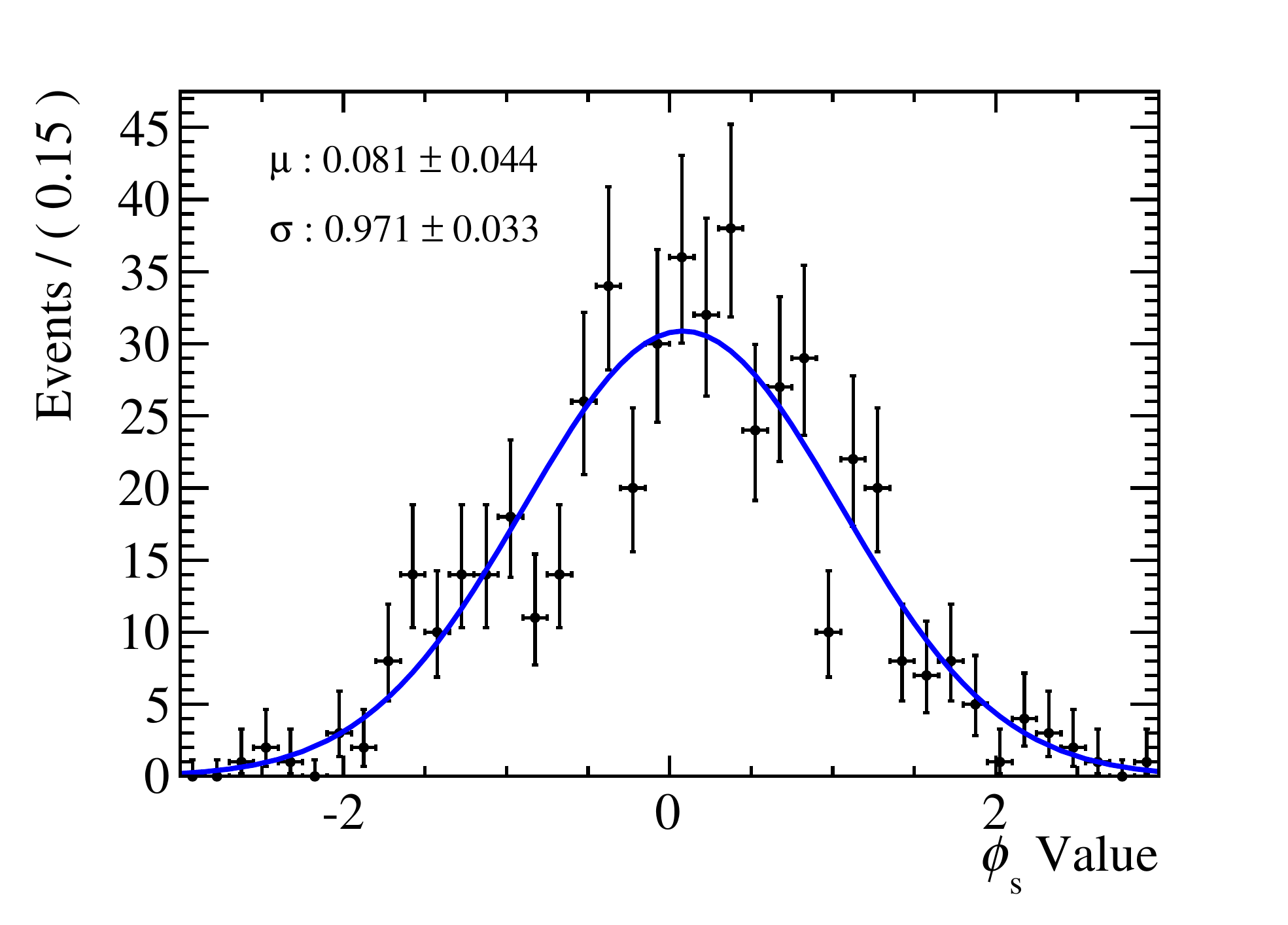}
\includegraphics*[width=0.38\textwidth]{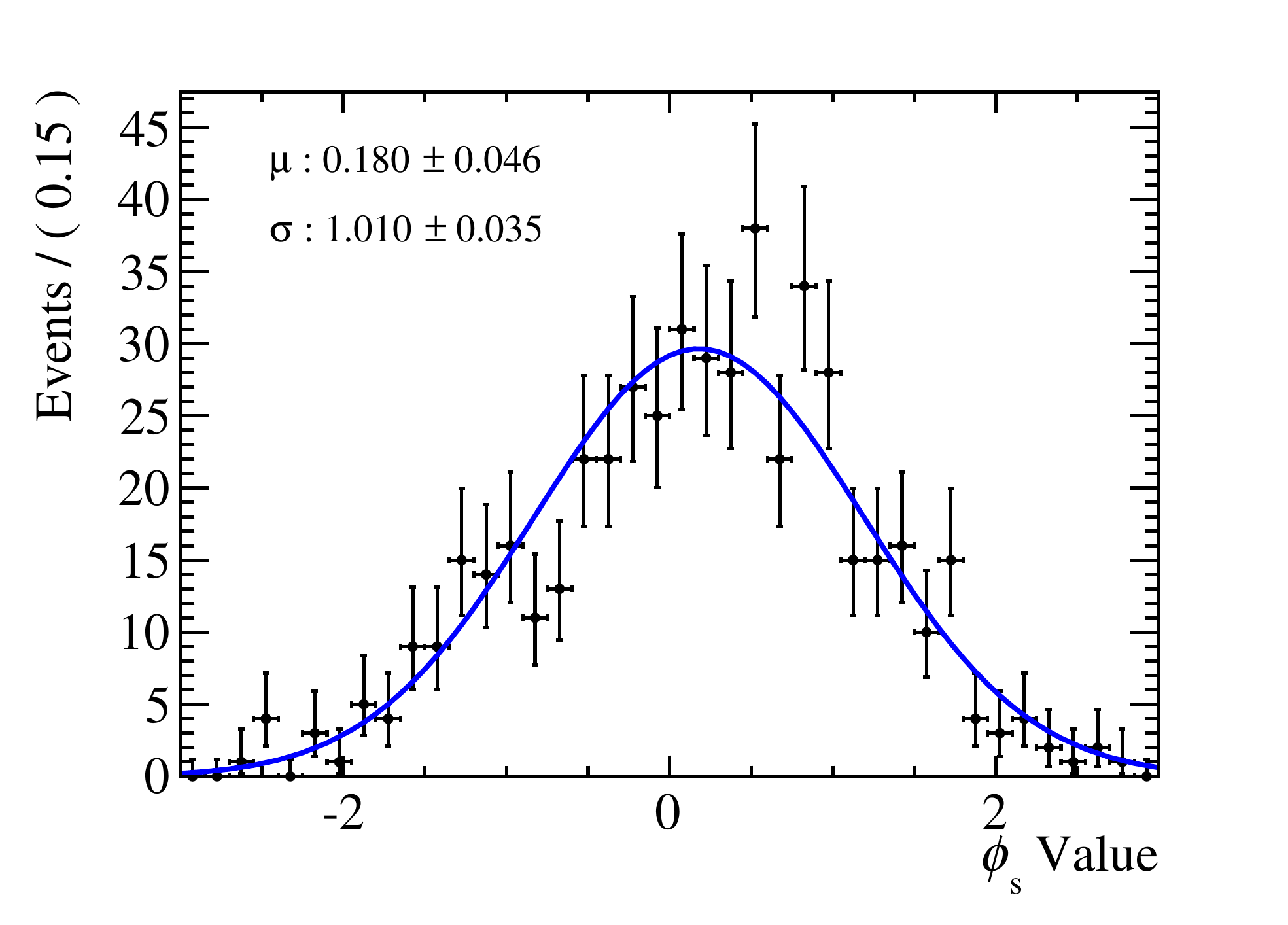}
\includegraphics*[width=0.38\textwidth]{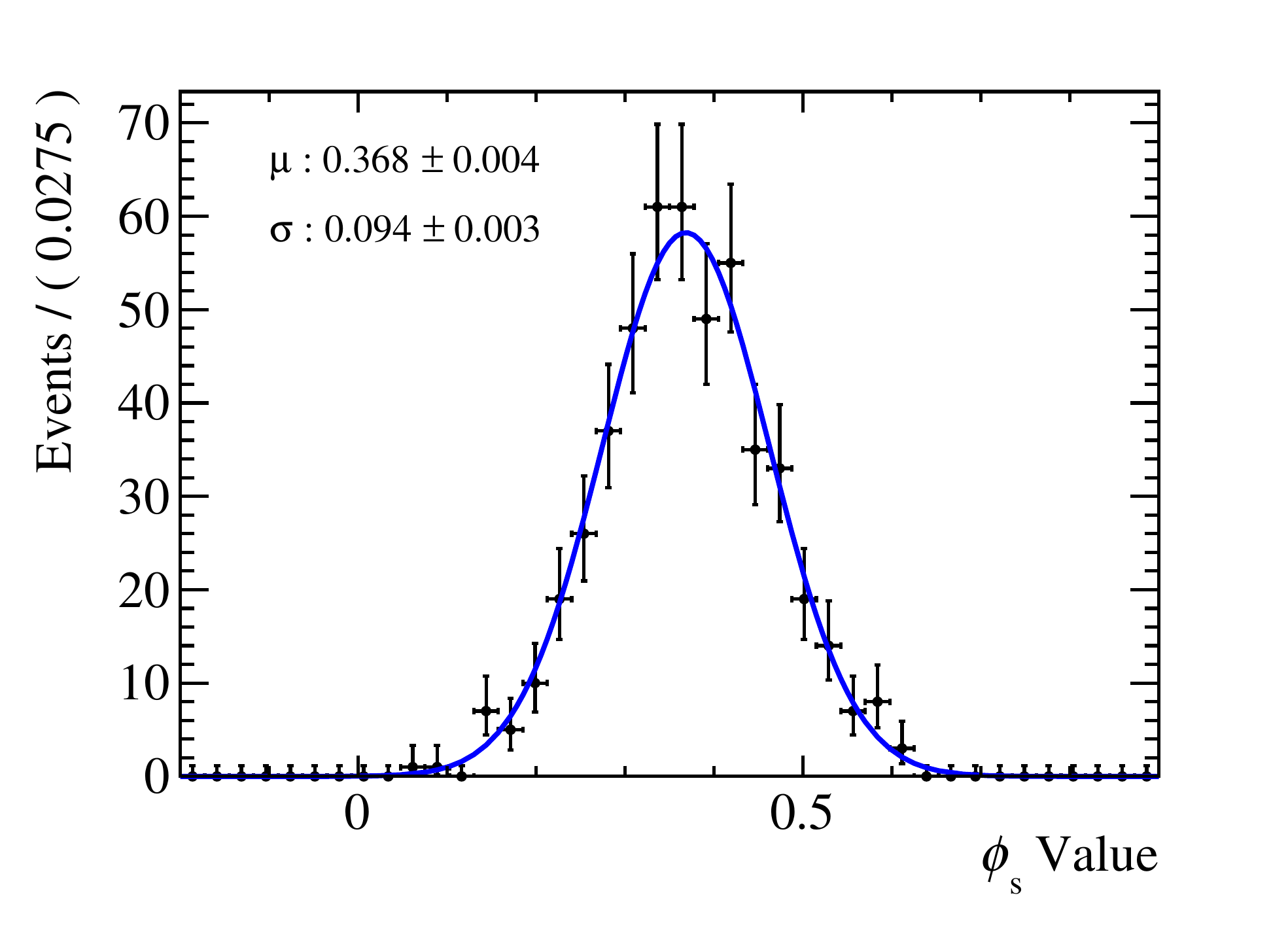}
\includegraphics*[width=0.38\textwidth]{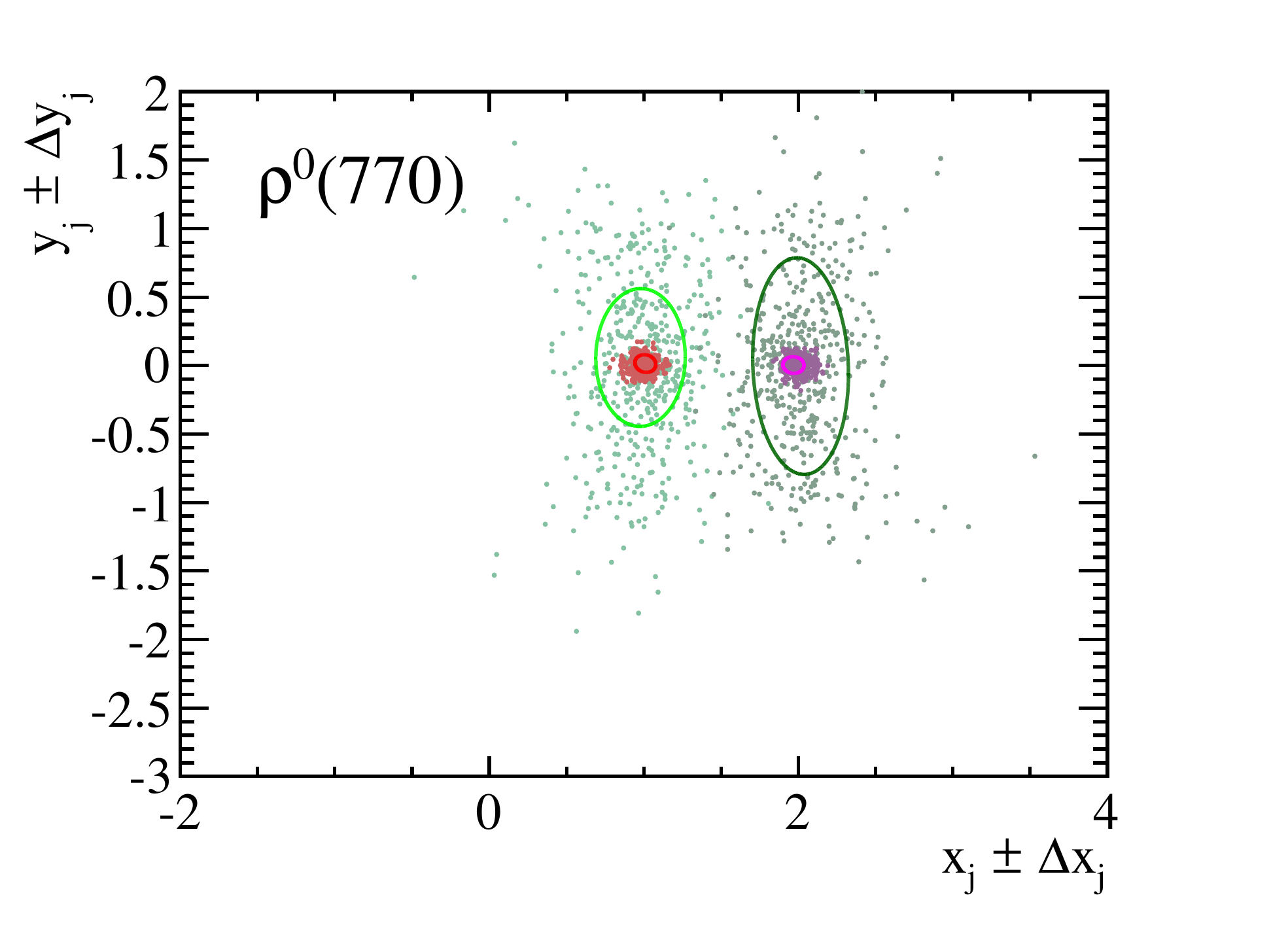}
\caption{
Fitted values of $\phi_{s}$ for
(top left) method ii with $\phi_{s} = -2\beta_{s}$, 
(top right) method ii with $\phi_{s} = -20\beta_{s}$ and
(bottom left) method iii with $\phi_{s} = -20\beta_{s}$.
The (bottom right) Argand plot displays the fitted coefficient values for the $\rho^{0}(770)$ resonance with fixed $\phi_{s} = -2\beta_{s}$ and ${\cal{A}}_{\CP} = 50\%$.
}
\label{fig:PhiS}
\end{figure*}

A further study is performed to investigate the sensitivity to the $\phi_{s}$ observable. 
Figure~\ref{fig:PhiS} compares the results from methods ii and iii (such a determination is not possible with method i).
It is clear that it is possible to determine the weak phase with both improved precision and greater accuracy when tagging is applied.
With perfect tagging, the precision on $\phi_{s}$ shows an order of magnitude improvement.
Using a more realistic tagging power of $\sim 5\%$, as achieved recently by LHCb~\cite{LHCb-PAPER-2014-038,LHCb-PAPER-2014-051}, still provides a factor $\sim 2.5$ better sensitivity to $\phi_{s}$ than the untagged case.
Alternatively one can fix the value of $\phi_{s} = -2\beta_{s}$ in the fit and float the $\Delta y_{j}$ parameter of the $\rho^{0}(770)$ resonance in order to measure the relative phase between the \Bs and \Bsb decay to this state. 
This approach is also illustrated in Fig.~\ref{fig:PhiS} and shows the same behaviour comparing methods ii and iii.

\section{Summary}
\label{sec:summary}

The recent observation of \Bs decays to charmless three-body final states 
marks the start of a new and interesting field of \CP violation investigation.
In this note, a comparative sensitivity study for different approaches to Dalitz plot analysis has been performed for \BsToKSpipi decays. 
It has been demonstrated that good precision for the phase difference
between \Bs and \Bsb decays to $K^{*\pm}(892)\pi^{\mp}$ can be achieved
with untagged analysis approaches (e.g. for the LHC Run I and II).
Flavour tagging is, however, needed to determine $\phi_{s}$ (i.e. the relative phase 
in $\Bs(\Bsb)\to\KS\rho^{0}(770)$ decays).
These results indicate directions for possible amplitude analyses that can be pursued in future by Belle~II and  LHCb.

\section*{Acknowledgements}

This work is funded by the European Research Council under FP7 and by the United Kingdom's Science and Technology Facilities Council.
The authors thank their colleagues in the LHCb collaboration for the experimental progress that provides the context for this work.

\newboolean{articletitles}
\setboolean{articletitles}{false}
\nocite{*}
\bibliographystyle{elsarticle-num-new}
\bibliography{references}

\end{document}